\documentclass[letter]{aa}
\usepackage{graphicx}
\usepackage{txfonts}
\begin{document}
   \title{Candidate tidal disruption events from the \\
   XMM-Newton Slew Survey}

    \author{P. Esquej
          \inst{1}\fnmsep\inst{2},
	  R.D. Saxton\inst{2},
	  M.J. Freyberg\inst{1},
	  A.M. Read\inst{3},      
	 B. Altieri\inst{2},
	 M. Sanchez-Portal\inst{2}	 
	 \and
	  G. Hasinger\inst{1}
         }

   \offprints{P. Esquej}

   \institute{
	Max-Planck-Institut f\"{u}r extraterrestrische Physik, Giessenbachstrasse, 85748 Garching, Germany\\
             \email{pili@mpe.mpg.de}
	     \and
	European Space Agency (ESA), European Space Astronomy Centre, Villafranca, Apartado 50727,  28080 Madrid, Spain
	\and
	Dept. of Physics and Astronomy, Leicester University, Leicester LE1 7RH, U.K.\\
             }

   \date{Received July 20, 2006; accepted December 11, 2006}

 
  \abstract
   {In recent years, giant amplitude X-ray flares have been observed from a handful of non-active galaxies. The most plausible scenario of these unusual phenomena is tidal disruption of a star by a quiescent supermassive black hole at the centre of the galaxy.}
   {Only a small number of these type of events have been observed and confirmed to date. The discovery of more cases would allow a number of fundamental conclusions to be drawn about properties such as the frequency of tidal disruption events, the distribution of quiescent black hole masses and their influence in the context of galaxy/AGN formation and evolution among others.}
   {Comparing the XMM-Newton Slew Survey Source Catalogue with the ROSAT PSPC All-Sky Survey five galaxies have been detected a factor of up to 88 brighter in XMM-Newton with respect to ROSAT PSPC upper limits and presenting a soft X-ray colour. X-ray luminosities of these sources derived from slew observations have been found in the range $10^{41}$ $-$ $10^{44}$ erg s$^{-1}$, fully consistent with the tidal disruption model. This model predicts that during the peak of the outburst, flares reach X-ray luminosities up to $10^{45}$ erg s$^{-1}$, which is close to the Eddington luminosity of the black hole, and afterwards a decay of the flux on a time scale of months to years is expected. Multi-wavelength follow-up observations have been performed on these highly variable objects in order to disentangle their nature and to investigate their dynamical evolution.}
   {Here we present sources coming from the XMM-Newton Slew Survey that could fit in the paradigm of tidal disruption events. X-ray and optical observations revealed that two of these objects are in full agreement with that scenario and three other sources that, showing signs of optical activity, need further investigation within the transient galactic nuclei phenomena.}
   {}

   
   \authorrunning{P. Esquej et al.}
   \titlerunning{Candidate tidal disruption events from the XMM-Newton Slew Survey
   }
   \maketitle


\section{Introduction}

The existence of supermassive black holes (SMBH) at the nuclei of active galaxies (AGN) is amply demonstrated by X-ray observations. This fact is accepted due to the detection of luminous hard power-law like X-ray emission, rapid variability and relativistic effects in the iron-K line profile. Accretion of gas onto these black holes as a mechanism to power AGN is supported observationally. Observations of other feeding processes in this context such as black hole mergers and tidal disruption of stars are still rare. There is now strong evidence that the centres of non-active galaxies are also occupied by concentrated dark objects, a phenomenon predicted long ago by theory (\cite{Lidskii}; \cite{Rees}). Proof of this is the discovery of giant-amplitude, non-recurrent X-ray flares observed from several non-active galaxies. Such phenomena are favorably explained in terms of tidal disruption of stars by a central supermassive black hole. Some debris would be expelled at high velocity when the star approaches the dark object and the remnant would be accreted. These events would generate giant-amplitude outbursts of X-ray radiation decaying on a time scale of several years.

Actually, all kind of high-amplitude variability events are of great interest in astrophysics as they often trace the footprints of extreme violent physical processes, which are features of cataclysmic mechanisms of accretion. Large-area X-ray surveys with high sensitivities and good spatial resolution are essential for studying the long-term temporal properties of the bulk of X-ray sources. Within the tidal disruption scenario, dramatic X-ray outbursts from several optically non-active galaxies have been discovered by ROSAT (\cite{Bade}; \cite{KomBade}) in the PSPC All-Sky Survey (RASS) and confirmed by XMM-Newton and Chandra (\cite{Kom04}). All of these galaxies present similar properties (\cite{Kom02}): extreme X-ray softness in outburst, huge X-ray peak luminosity up to $\sim10^{45}$ erg~s$^{-1}$, giant amplitude variability and absence of Seyfert activity.

\begin{table*}[ht]
  
\caption{Characteristics of the slew sources detected as candidate tidal disruption events ordered in decreasing flux ratio.}             
\label{table1}      
\centering                         
\begin{tabular}{c c c c c}        

\hline\hline                 
\textbf{Slew source} & \textbf{NED counterpart} & \textbf{Observation date} & \textbf{Count-rate$_{0.2-2keV}$} & \textbf{Flux ratio} \\ 
& & &  [cts s$^{-1}$] & EPIC-pn/PSPC-$2\sigma$-upper-limit\\
\hline                        

   XMMSL1 J111527.3+180638 & NGC 3599 & 2003$-$11$-$22 & 4.95 & 88 \\     
   XMMSL1 J132342.3+482701 & SDSS J132341.97+482701.3 & 2003$-$12$-$01 & 1.60 & 83 \\ 
   XMMSL1 J093922.5+370945 & SDSS J093922.90+370944.0 & 2004$-$05$-$18 & 1.90 & 81 \\
   XMMSL1 J020303.1$-$074154 & 2MASX J02030314$-$0741514 & 2004$-$01$-$14 & 2.16 & 63 \\
   XMMSL1 J024916.6$-$041244 & 2MASX J02491731$-$0412521 & 2004$-$07$-$25 & 1.89 & 21 \\
\hline   
                                
\end{tabular}
\end{table*}

\begin{figure*}[ht]
   \centering
   \resizebox{\hsize}{!}{\rotatebox[]{0}{\includegraphics[clip=true]{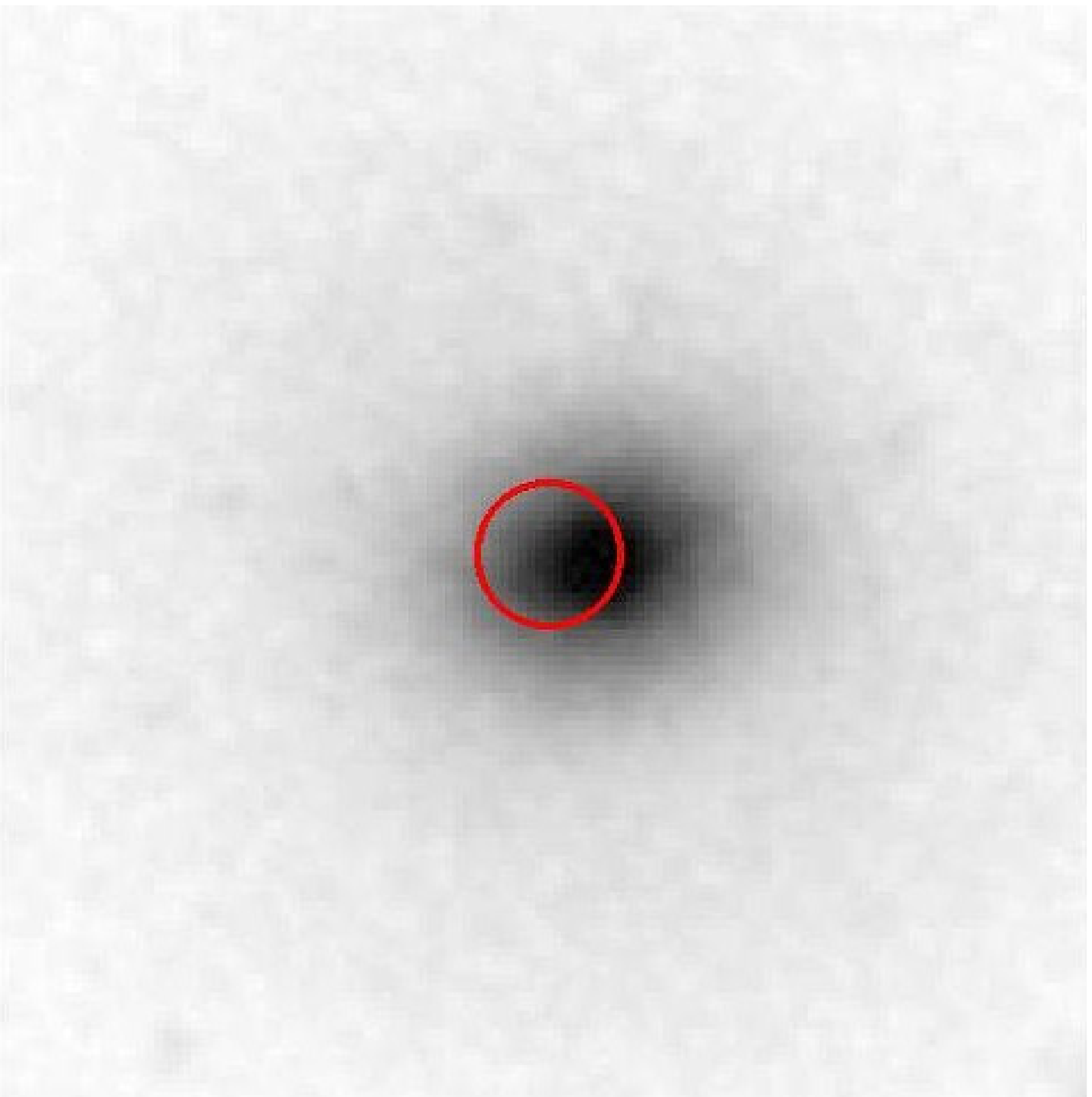}}
   \rotatebox[]{0}{\includegraphics[clip=true]{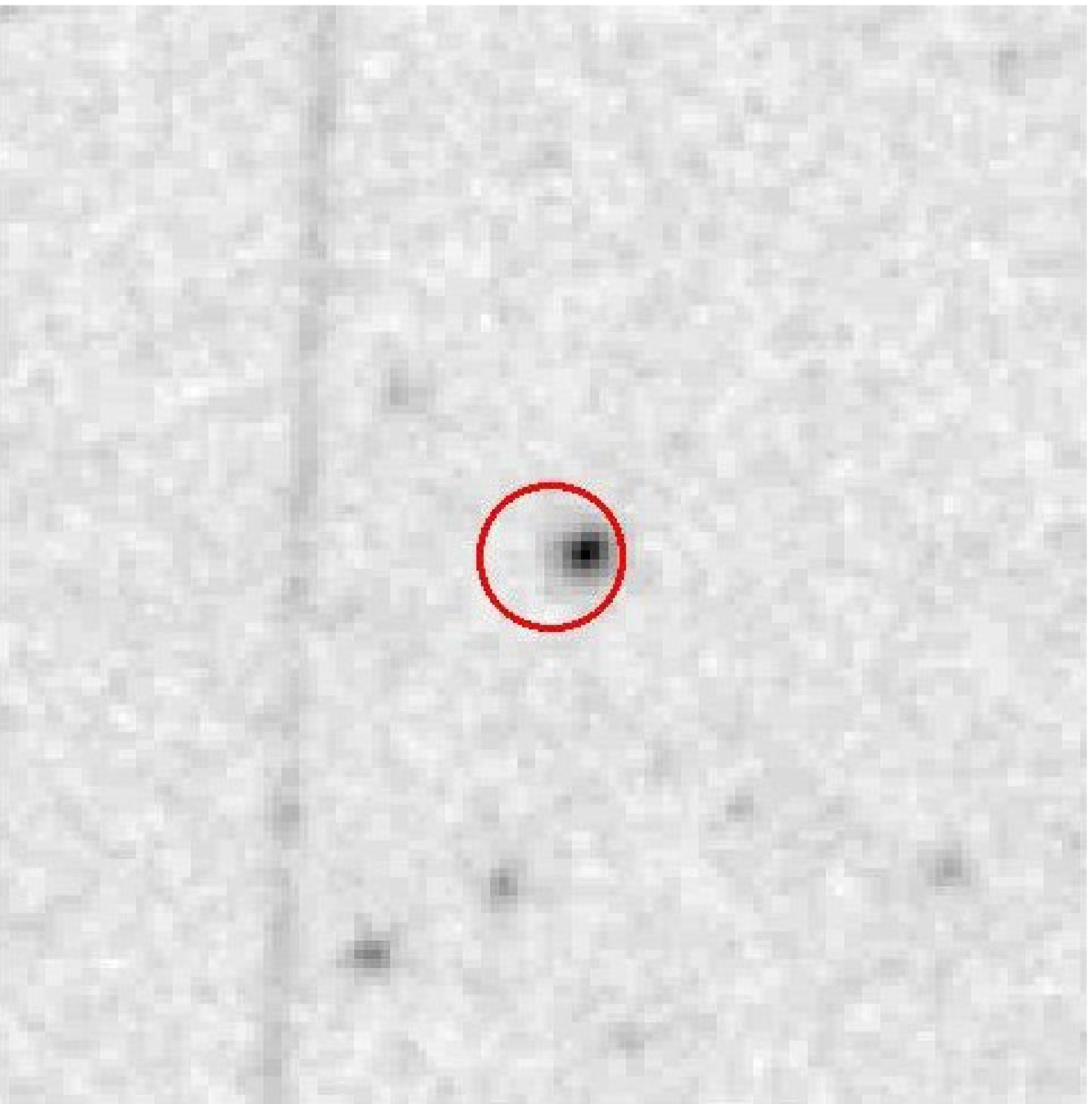}}   
   \rotatebox[]{0}{\includegraphics[clip=true]{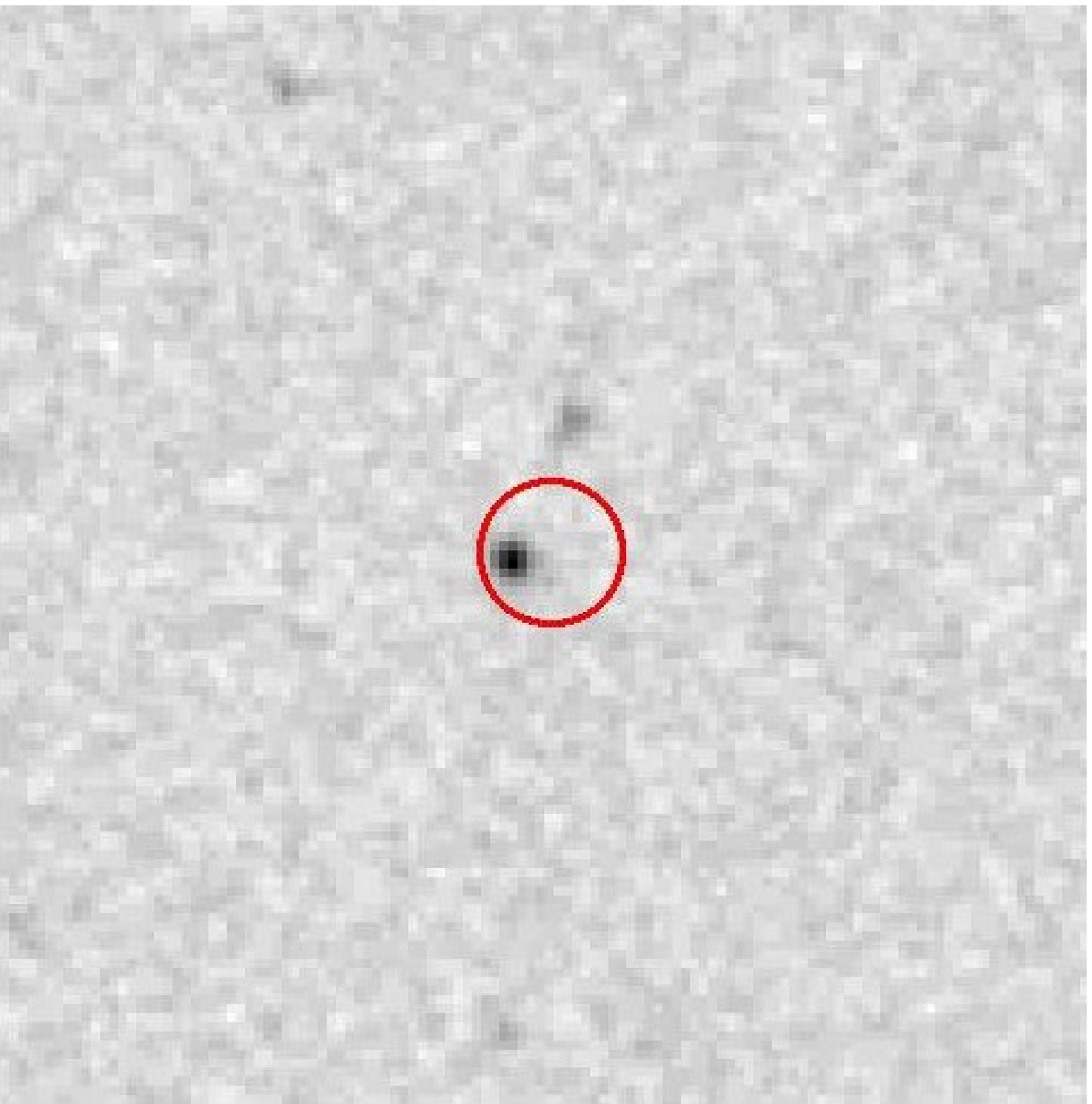}}
   \rotatebox[]{0}{\includegraphics[clip=true]{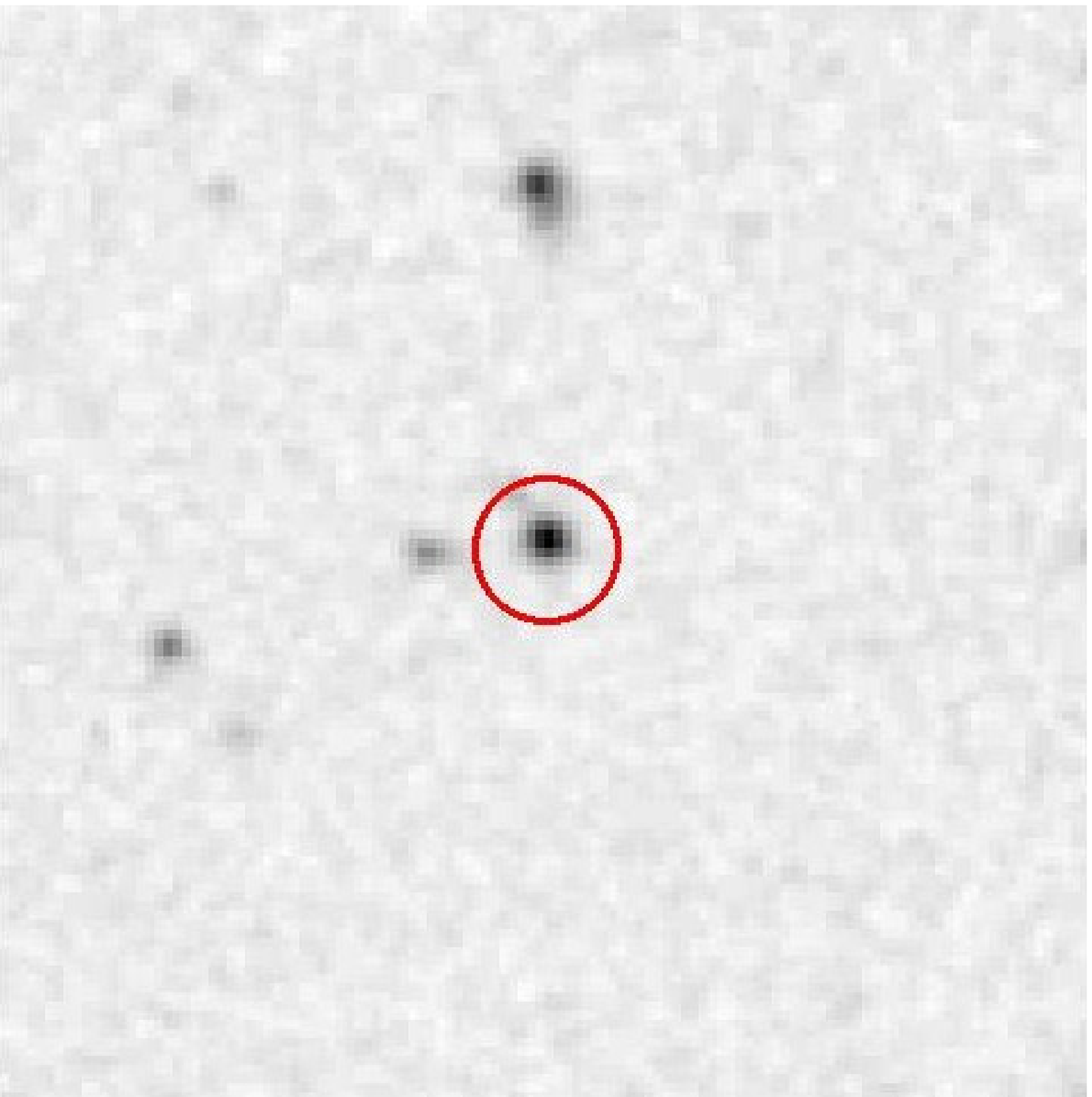}}
   \rotatebox[]{0}{\includegraphics[clip=true]{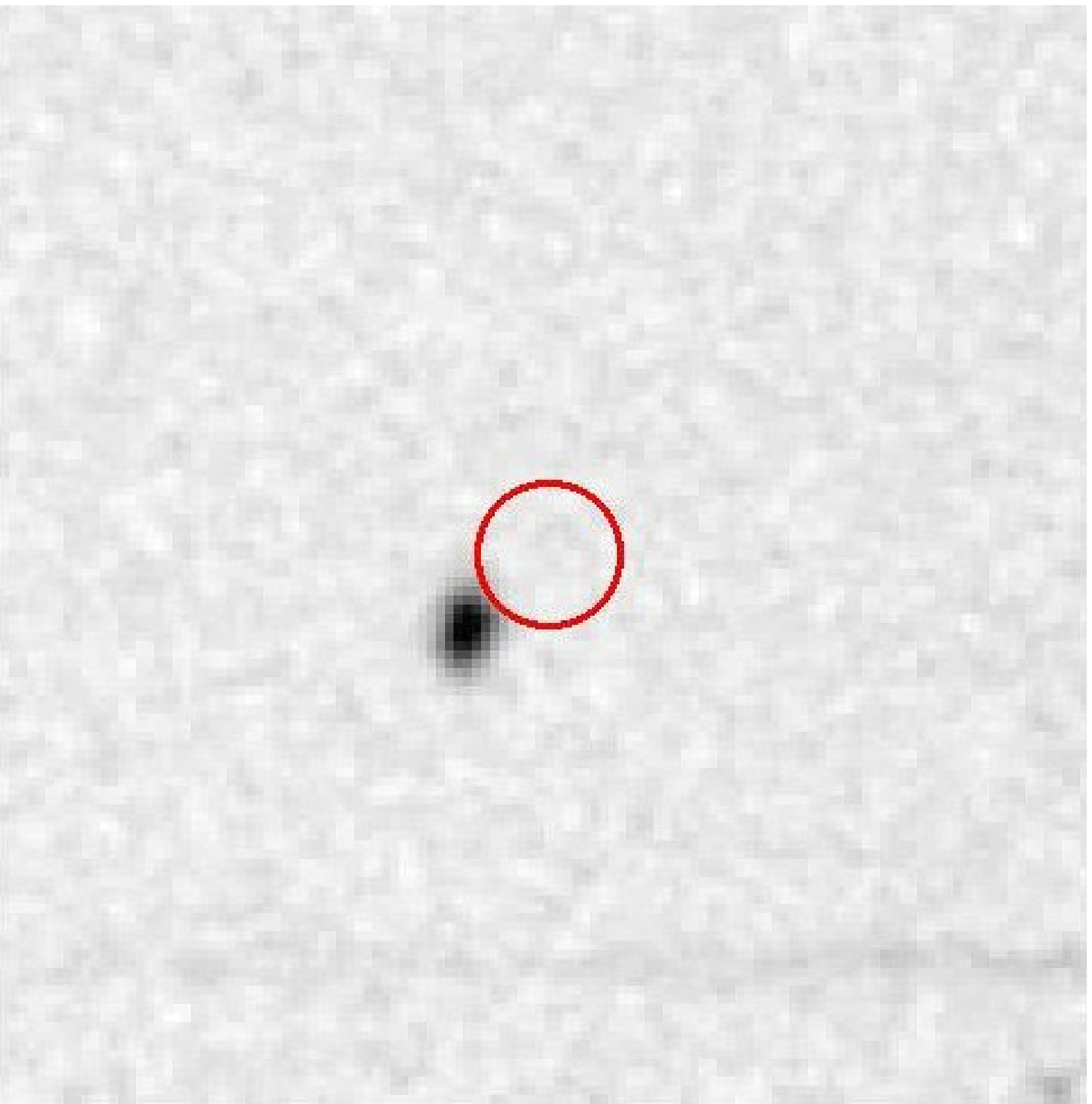}}}
   \put(-520,94){\tiny{XMMSL1 J111527.3+180638}}
   \put(-415,94){\tiny{XMMSL1 J132342.3+482701}}
   \put(-309,94){\tiny{XMMSL1 J093922.5+370945}}
   \put(-206,94){\tiny{XMMSL1 J020303.1-074154}}
   \put(-101,94){\tiny{XMMSL1 J024916.6-041244}}

     \caption{Positions of slew sources (red circles) presented in Table~\ref{table1} overplotted on DSS images of side 2 arcmin. The radius of the circles has been fixed to 8 arcsec, which is the astrometric uncertainty of the slew sources. Offsets between 2 and 6 arcsec have been found for all galaxy counterparts except a 12 arcsec offset for 2MASX~J02491731$-$0412521, whose identification is unambiguous as no other source lies nearby.
               }  
        \label{images}
    \end{figure*}

Here we report on highly variable objects detected in a high-state with XMM-Newton during slew observations. Systematic comparison of source fluxes determined from the slew survey with those in the RASS have been performed. These occurrences have been identified with EPIC-pn source soft-fluxes a factor between 21 and 88 brighter than their RASS PSPC upper limits. All these objects present a very soft X-ray colour and have a galaxy counterpart coincident with the XMM-Newton position. These characteristics allowed us to suggest these sources as tidal disruption candidates. 

AGN variability can not be ruled out \textit{a priori} for the sources discussed here because the possibility remains that a Seyfert nucleus could reside in these galaxies, undetected in their low-states by ground-based optical observations. Although the common X-ray variability factors found in AGN are $\sim$2$-$3, a handful of extreme variability cases with variability factors of up to several tens on time scales of a few years have also been detected. These are explained as sources whose spectrum changed appearance becoming reflection-dominated from Compton-thin or \textit{vice versa}, either due to a temporary switching-off of the nuclear radiation (\cite{Bianchi} and references therein) or a change in the line-of-sight absorbing column (\cite{Risaliti}). Many Narrow$-$line Seyfert~1 galaxies (NLS1) also show large soft X-ray variabilities, but on very short time scales (\cite{Boller}). Future planned X-ray observations of the sources presented in this paper will allow us to investigate the X-ray properties of the objects along with their temporal evolution.

\section{Candidates selection}\label{sec_obs}
The excellent capabilities of the XMM-Newton 
satellite (\cite{Jansen}) have been exploited for 
outstanding discoveries since its launch in December 1999. The great collecting area of its mirrors and the high quantum efficiency of the EPIC detectors confer on XMM-Newton the status of being the most sensitive X-ray observatory ever flown. This characteristic is strikingly apparent while maneuvering between observation targets. Yielding only at most 15 seconds of on-source time, sources detected during slew observations constitute a hard-band 2$-$12 keV survey from 5 to 10 times deeper than all other all-sky surveys and a soft-band 0.2$-$2 keV survey comparable with the RASS. A total of 219 XMM-Newton slew observations have been processed and source searched (\cite{Freyberg}, and references therein) giving a sky coverage of roughly 6300 square degrees ($\sim15\%$ of the sky). Of order of 4000 sources have been detected comprising the XMM-Newton Slew Survey Source Catalogue (XMMSL1) (Saxton et~al. in prep.), made public in May 2006.     
   
In order to identify cases of long-term X-ray variability, sources detected in XMM-Newton slew observations have been compared in terms of their fluxes in the soft 0.2$-$2.0 keV energy band with those lying in the overlapping fields of the RASS. From the flux ratio comparison highly variable sources emerged mainly corresponding to variable stars and AGN type objects. Optically unidentified sources have also been found with a high flux ratio that need follow-up observations to disentangle their nature. When no RASS source has been detected in the corresponding slew position, upper limits have been calculated. The ROSAT survey data have been analysed using the {\tt EXSAS} software package (\cite{Zimmermann}). $2\sigma$ upper limits have been determined with the command {\tt COMPUTE/UPPER\_LIMITS} around the XMMSL1 position in the 0.1$-$2.0 keV band with fixed source position and a source cut radius of 3 FWHM of the point spread function. For the comparison with XMMSL1 a threshold {\tt UL\_THR} of 8 has been selected. Corresponding $2\sigma$ upper limit count rates have been obtained by dividing by the vignetting corrected exposure map value.

Both EPIC-pn and PSPC count rates have been converted to 0.2$-$2.0 keV fluxes using PIMMS assuming a typical near-maximum tidal event spectrum (\cite{Kom02}). The Galactic $N_{H}$ has been derived for every source and a black body model with kT=0.07 has been applied. Five slew sources (Table~1; Fig.~\ref{images}) with an XMM-soft flux versus RASS $2\sigma$ upper limit ratio from 21 to 88, have been identified with galaxies for which no previous AGN activity was known. In addition, NGC~3599 lies in the field of view of two ROSAT PSPC pointed observations, 600263p and 300169p, allowing a more stringent statement about its high amplitude variability. The corresponding upper limits imply flux ratios of 297 and 100 respectively for this source (note that for the first observation
the source is located very close to the PSPC window support structure causing a higher systematic uncertainty).

\begin{table*}[ht]
\caption{Characteristics of the sources. Fluxes and luminosities have been derived from XMM-Newton slew observations.}             
\label{table2}      
\centering                          
\begin{tabular}{c c c c c}        
\hline\hline   
\textbf{Source name} & \textbf{Redshift}  & \textbf{F$_{0.2-2keV}$} & \textbf{L$_{0.2-2keV}$}  \\ 
& &[$erg$ $s^{-1}$ $cm^{-2}$] & [$erg$ $s^{-1}$]\\
\hline                        
   NGC 3599 & 0.0028 & $7.1\times10^{-12}$ & $1.22\times10^{41}$ \\ 
   SDSS J132341.97+482701.3 & 0.0875 & $2.3\times10^{-12}$ & $4.39\times10^{43}$\\    
   SDSS J093922.90+370944.0 & 0.1845 & $2.7\times10^{-12}$ & $2.63\times10^{44}$\\   
   2MASX J02030314$-$0741514 &  0.0615 & $3.1\times10^{-12}$ & $2.82\times10^{43}$\\
   2MASX J02491731$-$0412521 & 0.0186 & $2.7\times10^{-12}$ & $2.07\times10^{42}$\\
\hline                                   
\end{tabular}
\end{table*}

\section{Slew sources as candidates within the tidal disruption scenario}   
The best diagnostic to probe the existence of a dormant black hole at the centre of non-active galaxies 
is the detection of emerging flaring radiation produced when a star is tidally disrupted by the black hole (\cite{Rees}).  

To date, all five non-active galaxies supposed to contain a SMBH interpreted within this picture have been detected with ROSAT (\cite{Kom02} and references therein) and present these properties:
\begin{itemize}
\item giant-amplitude no-recurrent variability
\item enormous X-ray peak luminosity (up to $\sim~10^{45}$ erg s$^{-1}$ in maximum)
\item ultra-soft X-ray spectrum ($kT_{bb}\sim$0.004$-$0.1 keV applying a black body model)
\item no signs of AGN activity in ground-based optical spectra 
\end{itemize}
\noindent
with X-ray light curves consisting of a fast increase and a decline on a time scale from months to years, approximately following a $t^{-5/3}$ law. 

\begin{figure}[ht]
   \centering
   \resizebox{\hsize}{!}{\rotatebox[]{0}{\includegraphics[clip=true]{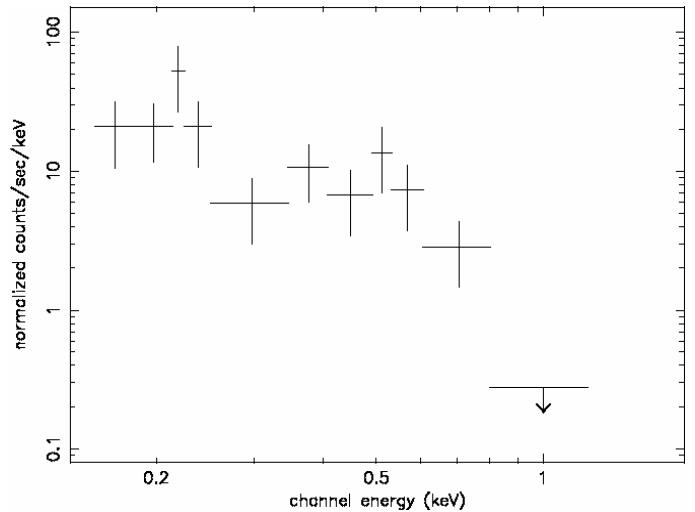}}} 
   \caption{EPIC-pn X-ray spectrum of NGC~3599. The arrow shows the upper limit above 0.8~keV.}
\label{spect1}
   \end{figure}

All five slew sources that could fit in this category have shown great variability in comparison with ROSAT upper limits (see Table~\ref{table1}). X-ray luminosities have been calculated (Table~\ref{table2}) assuming a $\Lambda$CDM cosmology with ($\Omega_{M}$,$\Omega_{\Lambda}$)~=~(0.3,0.7) and $H_{0}$~=~70~$km$~$s^{-1}$~Mpc$^{-1}$ and found to be between $10^{41}$ and $10^{44}$ erg s$^{-1}$, consistent with the tidal disruption model. These objects also are very soft (the source detection procedure in the XMM-Newton Slew survey has been performed independently in different energy bands (soft, hard and total band) and none of the sources have been detected in the hard 2$-$12 keV energy band). Although in most cases slew sources do not have enough photons to extract useful X-ray spectra, that of the slew counterpart of the galaxy NGC 3599 demonstrates that the source is very soft (Fig.~\ref{spect1}). 

\begin{figure}[ht]
\centering
   \resizebox{\hsize}{!}{\rotatebox[]{0}{\includegraphics[clip=true]{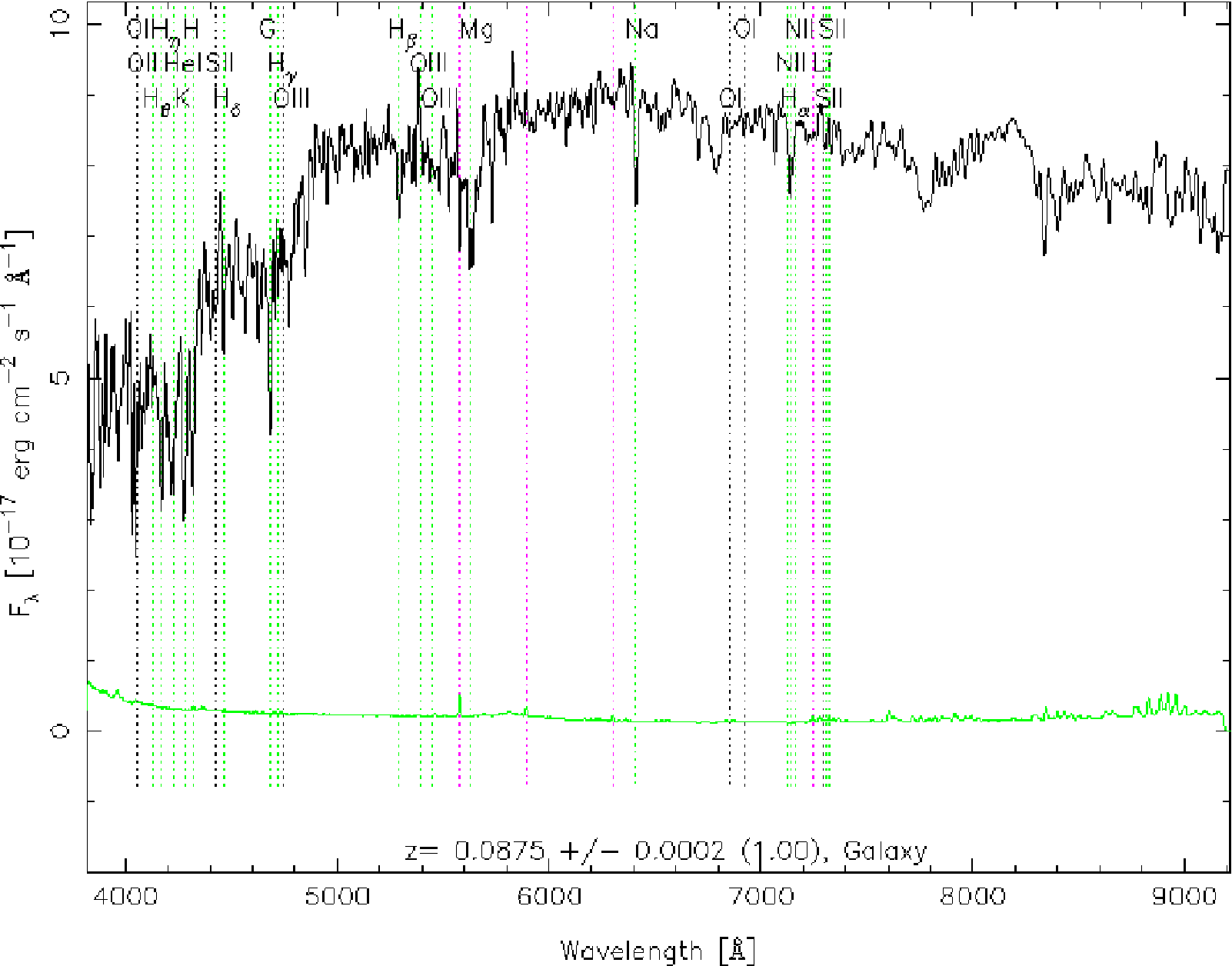}}}
    \caption{Optical spectrum of SDSS~J132341.97+482701.3 performed seven months before the corresponding slew exposure.}
\label{spect2}
   \end{figure}

NGC~3599 is classified as a non-active elliptical galaxy (\cite{Denicolo}) and SDSS~J132341.97+482701.3 is also non-active as its spectrum only shows absorption features from the atmospheres of individual stars (Fig.~\ref{spect2}). Therefore, both sources fulfill all properties for tidal disruption events.

The analysis of optical observations performed with the William Herschel Telescope (WHT) on the three remaining sources in the sample (their redshifts in Table~\ref{table2} have been derived from these optical spectra) reveals clear signatures of the existence of an AGN, even though in all three cases one or more diagnostics based on narrow emission lines (\cite{Veilleux}; \cite{Kauffmann}) do fail. Specifically, none of these objects shows [O~I]~6300~\AA\  emission, a good tracer of AGN activity (\cite{Baldwin}).

\begin{figure}[ht]
\centering
   \resizebox{\hsize}{!}{\rotatebox[]{0}{\includegraphics[clip=true]{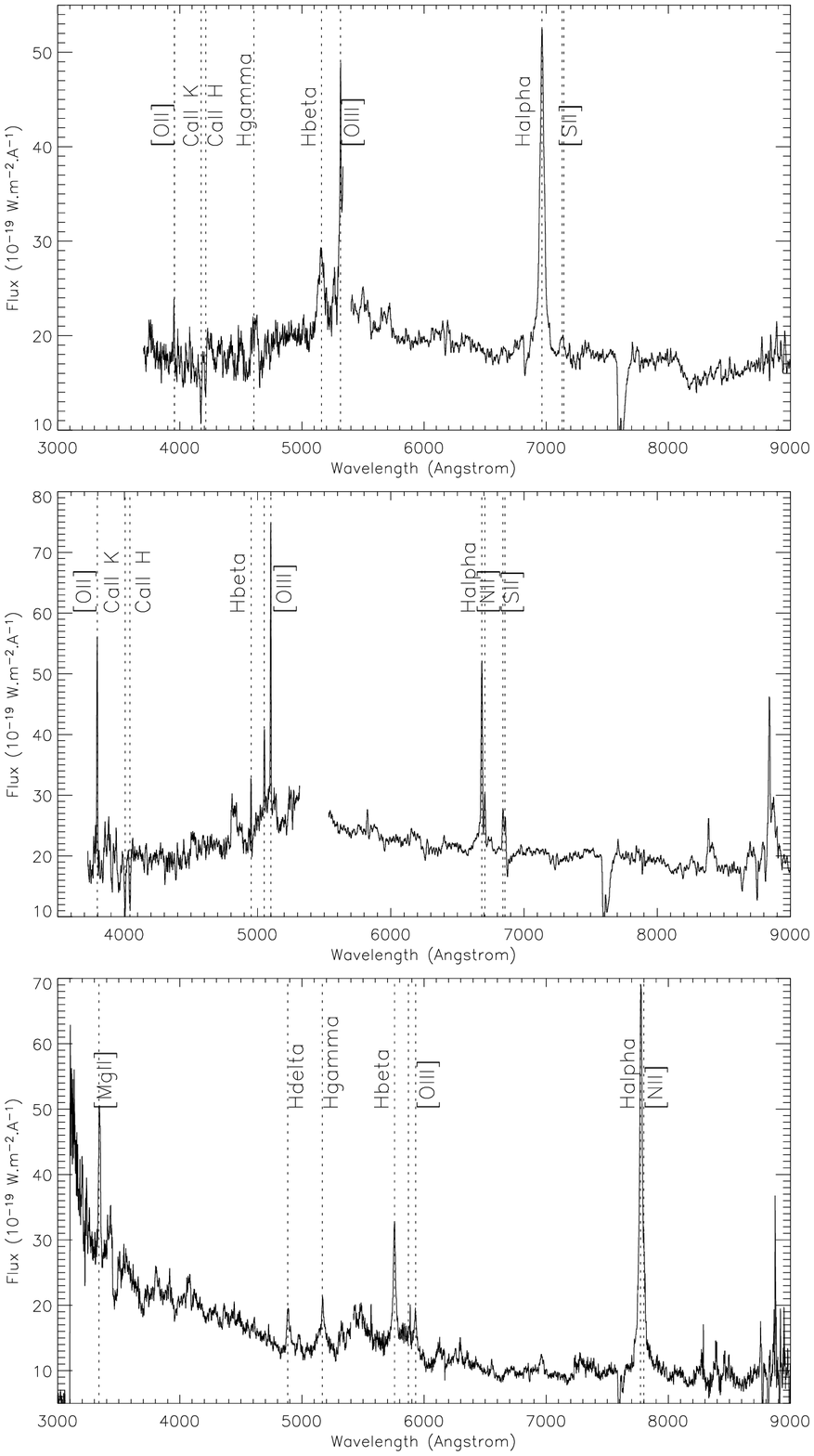}}}
    \caption{Optical spectra of 2MASX~J02030314$-$0741514, 2MASX~J02491731$-$0412521 and SDSS~J093922.90+370944.0 (from top to bottom).}
\label{spect3}
   \end{figure}

2MASX~J02030314$-$0741514 (Fig.~\ref{spect3}, top panel) could be classifed as a Seyfert~1 given the width and shape of the H$\beta$ line and the ratio [O~III]~5007~\AA/H$\beta$(total) $\sim$~3, although the [N~II] doublet is rather weak.

2MASX J02491731$-$0412521 (Fig.~\ref{spect3}, medium panel) presents narrow permitted emission lines, [O~III]~5007~\AA/H$\beta$ $>$3 and [S~II] (6716~,~6371~\AA)/H$\alpha$ $>$0.3. In addition, the observed broad base on the H$\alpha$ line leads to a Seyfert~1.9 classification (although this source also fails the [N~II]~6583~\AA/H$\alpha$ AGN diagnostic).

Finally, SDSS~J093922.90+370944.0 (Fig.~\ref{spect3}, bottom panel)~can be classified as a Narrow$-$line Seyfert~1 galaxy based on the width of the H$\alpha$ line ($\sim$ 950 km~s$^{-1}$), the strength of the Fe~II multiplets at 4570~\AA\ and 4924~\AA~and the ratio [O~III]~5007~\AA /H$\beta$(total) $\simeq$ 0.35 (\cite{Goodrich} and references therein). NLR diagnostics are unreliable due to the failure of producing a stable deblending of the H$\alpha$ line complex (narrow and broad components of the Balmer recombination line plus [N~II] doublet at 6548,~6583~\AA).

  \section{Conclusions}
Here we have described initial scientific results from the XMM-Newton Slew Survey for highly variable objects in the context of tidal disruption of a star by a supermassive dormant black hole (for a general view of the first scientific results of this survey see \cite{Read}). Although other scenarios have been considered to explain this kind of outburst (\cite{Kom02}) they have been rejected as they can not explain the observed properties of the sources presented here. It is worth noting that NGC~3599 was observed with an X-ray luminosity of $\sim$\,$10^{41}$ erg~s$^{-1}$, which is at the extreme upper end of ultra-luminous X-ray sources (ULX) observed in other galaxies, but an ULX interpretation is unlikely given the very high amplitude of variability detected in this source.

Five very soft sources hitherto classified as normal galaxies have been detected during slew observations showing high amplitude variability with respect to RASS PSPC upper limits. Such variability and other characteristics of these sources like their softness and luminosities, allowed us to propose them as tidal disruption events. Two of the sources presented here are in full agreement with the tidal disruption model as they are optically non-active galaxies and meet all known criteria for objects within this scenario. Planned follow-up observations on these targets will provide strong constraints on the physical context of the tidal disruption phenomenon. The three remaining sources show signs of AGN activity derived from optical observations and need to be investigated within the transient active galactic nuclei environment. Future X-ray observations are going to be performed on all five targets to further analyse their dynamical evolution.

The XMM-Newton Slew Survey when compared with ROSAT data will undoubtedly detect
further examples of tidal disruption events and other transient sources as the survey grows. The discovery and follow-up of each new case would begin to allow conclusions to be drawn about the feasibility and mechanisms involved in the phenomenon. We note that it would be of tremendous benefit for the slew data to be processed rapidly so that flaring events can be caught near their peak luminosity and follow-up observations scheduled in a timely manner. In addition, future X-ray surveys, like that planned with eROSITA (\cite{Predehl}), will be valuable in finding more of these outstanding sources.

\begin{acknowledgements}
We would like to thank S.~Komossa for her useful comments. The XMM-Newton project is an ESA Science Mission with instruments
and contributions directly funded by ESA Member States and the
USA (NASA). The XMM-Newton project is supported by the
Bundesministerium f\"ur Wirtschaft und Technologie/Deutsches Zentrum
f\"ur Luft- und Raumfahrt (BMWI/DLR, FKZ 50 OX 0001), the Max-Planck
Society and the Heidenhain-Stiftung. This research has made use of the NASA/IPAC Extragalactic Database (NED) and the NASA/IPAC Infrared Science Archive which are operated by the Jet Propulsion Laboratory, California Institute of Technology, under contract with the National Aeronautics and Space Administration. PE acknowledges support from the International Max Planck Research School on Astrophysics (IMPRS). 
\end{acknowledgements}

\end{document}